\documentclass[twocolumn,prl,groupedaddress,showpacs,superscriptaddress]{revtex4-2}
\usepackage[utf8]{inputenc}

\usepackage{hyperref}
\usepackage{hyperref}
\usepackage{booktabs}
\usepackage{upgreek} 

\usepackage{float}
\usepackage{graphicx}
\usepackage{epstopdf}
\usepackage{tikz-cd}
\usepackage{quiver}

\usepackage{amscd,amssymb,amsmath,latexsym,bm}
\usepackage{amsmath,amsthm,amssymb}
\usepackage{amsfonts}
\usepackage[mathcal,mathscr]{euscript}
\usepackage{braket}

\newcommand{\CM}{{\mathbb C}}

\newcommand{\ZM}{{\mathbb Z}}

\newcommand{\Dd}{{\mathcal D}}

\newcommand{\Hh}{{\mathcal H}}

\begin{document}

\title{A Universal Chern Model on Arbitrary Triangulations}

\author{Nigel Higson}
\email{higson@psu.edu}
\address{Department of Mathematics, Penn State University, University Park, PA 16802, USA.
}

\author{Emil Prodan}
\email{prodan@yu.edu}
\address{Department of Physics, Yeshiva University, New York, NY 10016, USA
}

\begin{abstract} Given a triangulation of a closed orientable surface, we place single-mode resonators or single-orbital artificial atoms at its vertices, edges and facets, and we devise near-neighbor hopping terms derived from the  boundary and Poincar\'e duality maps of the simplicial complex of the triangulation. Regardless of the surface or its triangulation, these terms always lead to tight-binding Hamiltonians with large and clean topological spectral gaps, carrying non-trivial Chern numbers in the limit of infinite refinement of the triangulation. We confirm this via numerical simulations, and demonstrate how these models enable topological edge modes at the surfaces of real-world objects. Furthermore, we describe a metamaterial whose dynamics reproduces that of the proposed model, thus bringing the topological metamaterials closer to real-world applications. 
\end{abstract}

\maketitle

The Haldane model \cite{HaldanePRL1988} is the prototypical example of a topological insulator from class A of the classification table of topological insulators and superconductors \cite{SRFL2008,QiPRB2008,Kit2009,RSFL2010}.
Indeed, any other A-class model on a uniform discretization of the plane can be adiabatically deformed into decoupled layers of Haldane Hamiltonians \cite{ProdanSpringer2016}. It will be useful to have a similar model on triangulations of generic compact orientable surfaces. In such settings, the infinite-system limit can be achieved by sequential refinement of the triangulations \cite{Footnote1}, and it will be in this limit where Chern numbers and bulk-boundary correspondences make sense. However the topological effects are also expected for fine-enough finite triangulations.
 
\begin{figure}
    \centering   \includegraphics[width=0.8\linewidth]{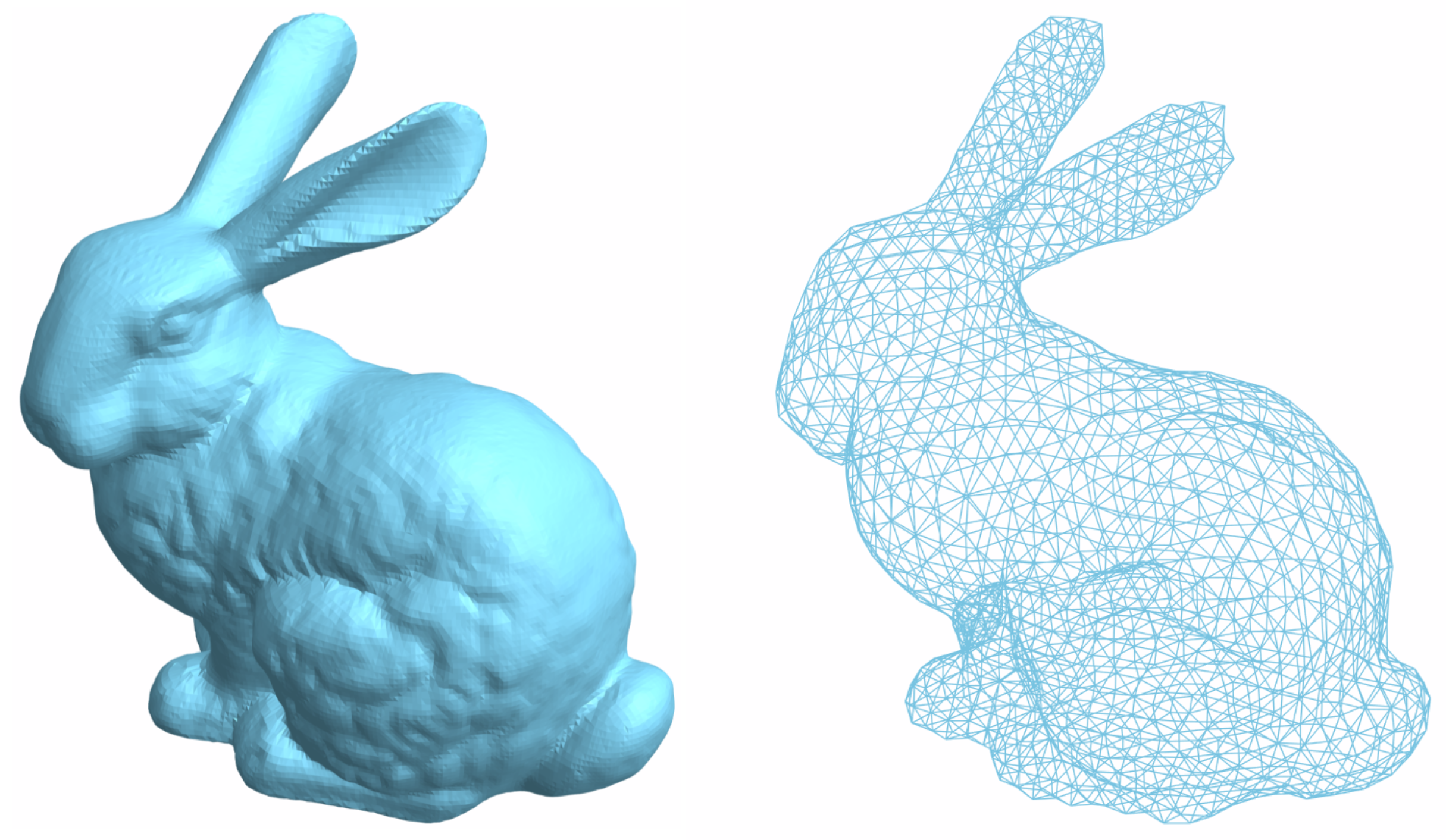}
    \caption{Left: Stanford Bunny, a triangulated mesh containing 69,451 faces generated by a 3D scan of a ceramic figurine \cite{Bunny}. Right: A simplified version containing 1549 vertices, 4641 edges and 3094 faces, to be used in our computer simulations.}
    \label{Fig:TE}
\end{figure}

 To convey the difficulty of such a task, we display in Fig.~\ref{Fig:TE} triangulations of the surface of a real-world object, highlighting the complexities and irregularities of such discrete geometries. Triangulations of surfaces of real-world objects are routinely generated by 3D scanners and, in fact, the stereolithography (stl) files inputted to 3D printers and computer numerical controlled (CNC) machines contain triangulations of surfaces. It is therefore natural to ask whether such triangulations can be decorated with suitably coupled resonators or artificial atoms ({\it e.g.} quantum dots) that exhibit topological collective dynamics, thus coating the real-world objects with defect-free topological classical or quantum metamaterials \cite{Footnote2} (see also Supplemental Material \cite{Supplemental}).
 
 Up to now, there has existed no algorithm that generates a clean Chern model on such discrete geometries. Even in the planar case, where topological gaps can be opened by applying physical or synthetic magnetic fields perpendicular to a material's surface, when the underlying   discrete geometry is irregular, the small gaps seen in the spectra \cite{MitchellNatPhys2018,BourneJPA2018} are not clean, but rather small mobility gaps contaminated with impurity states. Moreover, while the presence of a magnetic field can be theoretically implemented by twisting an ordinary model by the area 2-cocycle of the supporting surface \cite{BourneAHP2020}, a process that inserts the usual Peierls phase factors, this will be a very difficult task for geometries as in Fig.~\ref{Fig:TE}.

 The objective of this letter is to address this issue by supplying a general algorithm for a clean Chern model on an arbitrary triangulation. For a triangulated surface: 1) We give precise instructions as to where in space to place the resonators or the artificial atoms; 2) We specify the exact values of the pairwise coupling or hopping coefficients; 3) We give precise instructions on how to translate the abstract tight-binding models into (passive) phononic metamaterials with matching dynamics.
 
 The solution is based on the pure mathematical results from \cite{HigsonKTI2004,HigsonKTII2004,HigsonKTIII2004}, where an operator-theoretic generalization of surgery theory was developed. The central structures in classical surgery theory are  CW-complexes satisfying a form of Poincar\'e duality or, in short,  Poincar\'e complexes \cite{WallBook}. They were promoted in \cite{HigsonKTI2004} to complexes of Hilbert spaces that admit a Poincar\'e duality operator, now called Hilbert-Poincar\'e complexes. The boundary map and a self-adjoint version of the duality operator of such complex were assembled in \cite{HigsonKTI2004} into a pair of gapped operators over the total Hilbert space of the complex. This pair was used to define an analytic signature that is a homotopy invariant of the complex. 
 
 The analytic surgery theory of \cite{HigsonKTI2004,HigsonKTII2004,HigsonKTIII2004} is very general and context free (see \cite{Supplemental} for further details). As an application, Ref.~\cite{HigsonKTII2004} shows how to construct Hilbert-Poincar\'e complexes from triangulations, and this is what we will use here. In this context, the pair of gapped operators are local tight-binding operators that can be controlled in the infinite-lattice limit \cite{HigsonKTII2004}. Regardless of the surface or its triangulations, these limits  always fall in the same non-trivial stable homotopy classes of projections in the Roe $C^\ast$-algebra of the plane, the algebra where the operators belong after taking the infinite-lattice limit \cite{HigsonKTII2004}. These classes carry precisely the Chern numbers $\pm 1$, a fact that can be verified computationally. 
 
 One goal of this letter is to distill this construction down to the level of an actual meta-material and to demonstrate through numerical simulations the remarkable properties of the so-derived Hamiltonians. In the process, we confirm that the algorithm can be deployed on finite triangulations of real-world objects, where it produces clean topological gapped spectra, as predicted by \cite{HigsonKTI2004,HigsonKTII2004,HigsonKTIII2004}. Furthermore, we confirm that, on such finite triangulations, one-way topological wave-channels \cite{Footnote3} can be generated on demand by simply jamming the dynamics in the regions outside of the wave-channels. A Mathematica\textsuperscript \textregistered\ script generating the desired couplings for any given triangulation is supplied in \cite{Supplemental}.

 \begin{figure}
    \centering   \includegraphics[width=1.0\linewidth]{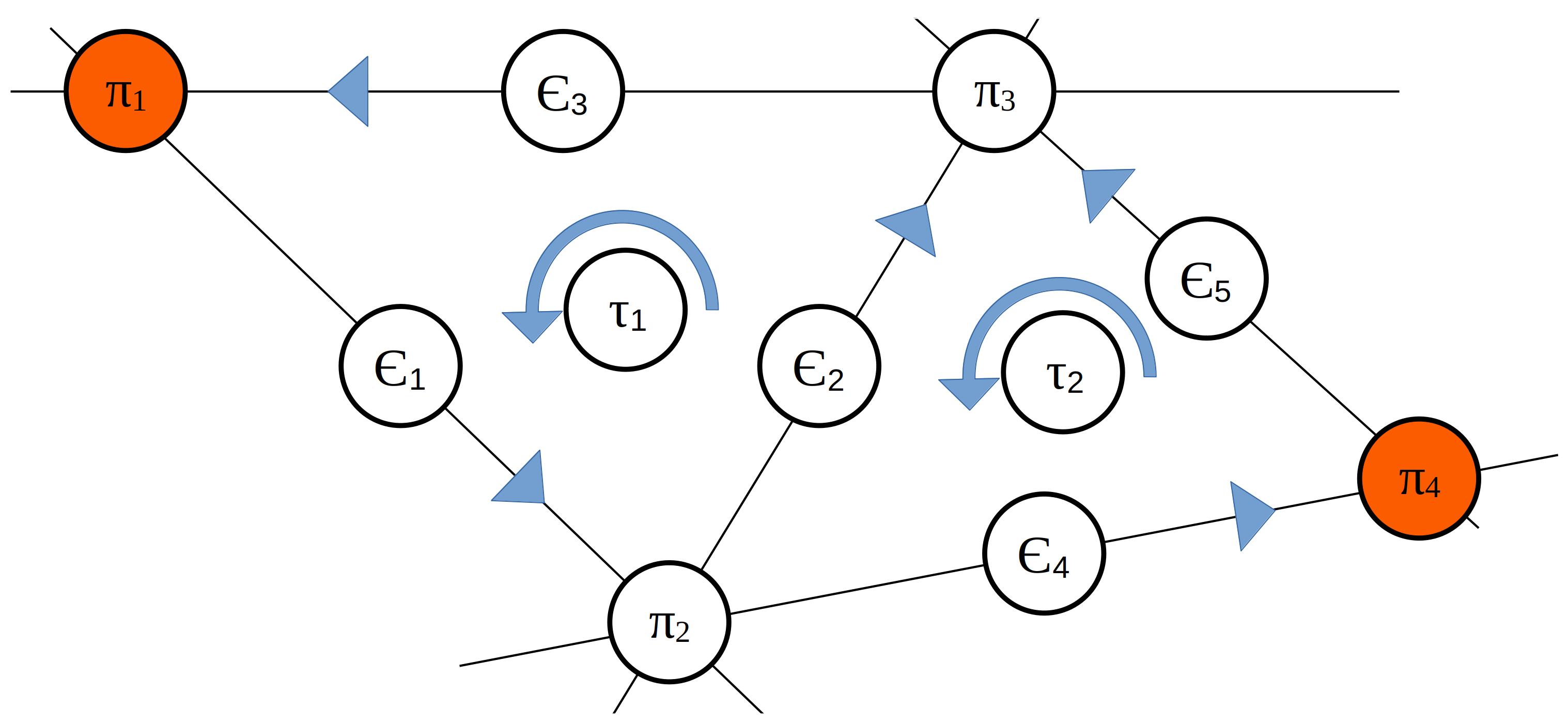}
    \caption{A region of a triangulation and its associated data, consisting of labels and orientations of the simplices. The marked vertices and the seen orientations order the vertices of the faces as $|\tau_1\rangle = |\textcolor{red}{\pi_1}\pi_2\pi_3\rangle$ and $|\tau_2\rangle = |\textcolor{red}{\pi_4}\pi_3 \pi_2 \rangle$. The triangulation is populated with single-mode resonators (the disks) in a one-to-one fashion with its simplices.}
    \label{Fig:TM}
\end{figure}

{\it From triangulations to topological materials.} A triangulation of a closed orientable surface $X$ supplies a finite simplicial complex, made up of $2$-simplices, the faces of the triangles ($\tau$), $1$-simplices, the edges ($\epsilon$), and $0$-simplices, the vertices ($\pi$). A $p$-simplex is specified by the set $[v_0,\ldots,v_p]$ of its vertices. Since the surface is orientable, the 2-simplices have preferential orientions. Additionally, the 2-simplices will be equipped with full orders of their vertices. The edges will also be equipped with orientations, but that will rather be arbitrary.  Specialized computer software, such as Mathematica\textsuperscript \textregistered\   automatically supplies a predefined order and orientation for the simplices extracted from a triangulation. That fits very well with our purposes.

We place identical single-mode resonators or single-orbital artificial atoms at the center of mass of the simplices, that is, at the vertices, at the middle of the edges, and at the center of mass of the triangles, as exemplified in Fig~\ref{Fig:TM}. If $|\bm x\rangle$ is the mode or orbital carried by the resonator or artificial atom positioned at $\bm x$, then the Hilbert space $\Hh$ where the collective dynamics happens is spanned by $|\bm x\rangle$, where $\bm x$ samples the center of mass of all simplices. However, it will be very inefficient to describe the couplings with this notation. Since each geometric position $\bm x$ is determined by a simplex, we will use the simplex instead (see Fig~\ref{Fig:TM} for examples). Simplices have differing dimensions, hence the Hilbert space splits as $\Hh = \bigoplus_{p=0}^2 \Hh_p$. For the state carried by a $p$-simplex $\xi=[v_0,\ldots,v_p]$, one enforces the convention 
\begin{equation}
\label{eq-antisymmetrization-condition}
    |\xi\rangle = |[v_0,\ldots,v_p] \rangle = (-1)^\sigma |[v_{\sigma(0)} \ldots v_{\sigma(p)}]\rangle,
\end{equation}
where $\sigma$ is any permutation and $v_j$'s are assumed to respect the predefined orientation of the simplex. For example, for $\tau_1$ from Fig.~\ref{Fig:TM}, we can take $v_0 {=} \pi_1$, $v_1{=}\pi_2$ and $v_2 {=} \pi_3$. On the other hand, if we choose $v_0 {=} \pi_3$, $v_1{=}\pi_2$ and $v_2 {=} \pi_1$, then $|[v_0,v_1,v_2]\rangle= - |\tau_1\rangle$. At one point, we will need to present a simplex as a fully ordered sequence of vertices, in which case we write its mode as $|v_0,\ldots,v_p\rangle$.

\begin{figure}
    \centering   \includegraphics[width=1.0\linewidth]{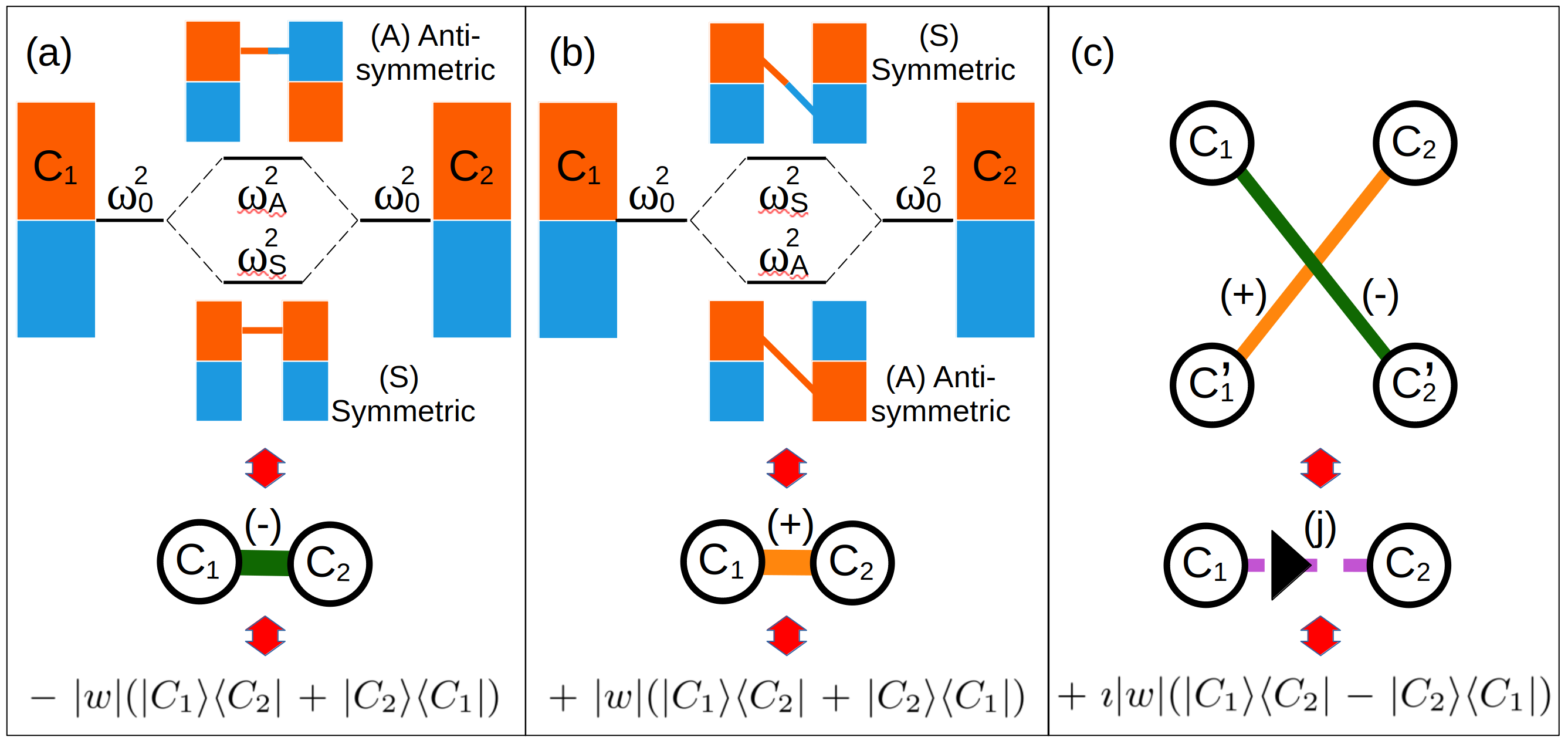}
    \caption{Implementation of tight-binding terms using coupled acoustic cavities (C) carrying resonant modes $|C\rangle$ with fundamental frequency $\omega_0$. Each panel shows the physical connections (top), their symbols (middle) and the hopping terms (bottom) they implement. Panel (c) reduces a purely imaginary hopping term to the connections from panels (a) and (b), at the expense of introducing copies of the resonators (see main text). The value of $w = \tfrac{1}{2}(\omega_S^2 - \omega_A^2)$ can be adjusted by varying the cross section of the bridges. The coloring of the cavities is a rough representation of the pressure field amplitudes at the corresponding resonant frequencies.}
    \label{Fig:CR}
\end{figure}

The proposed Hamiltonians take the tight-binding form $\sum w_{\xi,\xi'} |\xi\rangle \langle \xi' |$, where the sum is over pairs of simplices from the triangulation. Metamaterials science is a mature field and transformation of abstract tight-binding Hamiltonians into actual metamaterials with matching dynamics is now routinely done via the coupled-mode theory \cite{ChenSB2025}. As an example, Fig.~\ref{Fig:CR}(a,b) shows two identical acoustic cavities (C$_{1,2}$) with colorings depicting the amplitudes of air pressure fields corresponding to their first resonant modes $|C_1\rangle$ and $|C_2\rangle$ \cite{Supplemental4}. The two cavities are coupled via short air channels (aka tubes). As a result, the resonant modes hybridize and the resonant pressure fields  and frequencies of the coupled dimmers are as presented in Fig.~\ref{Fig:CR}(a,b), when the air channels connect different points of the cavities. The dynamics of the seen hybridizations are accurately described by the hopping terms listed at the bottom of the diagrams \cite{Supplemental}. Since the strength of the hoppings can be varied by adjusting the cross sections of the air channels, one can translate any positive and negative hopping terms of an abstract tight-binding Hamiltonian into actual acoustic couplings. The implementation of the complex hoppings in Fig.~\ref{Fig:CR}(c) follows \cite{BarlasPRB2018} and is discussed below.

\begin{figure}
    \centering   \includegraphics[width=1.0\linewidth]{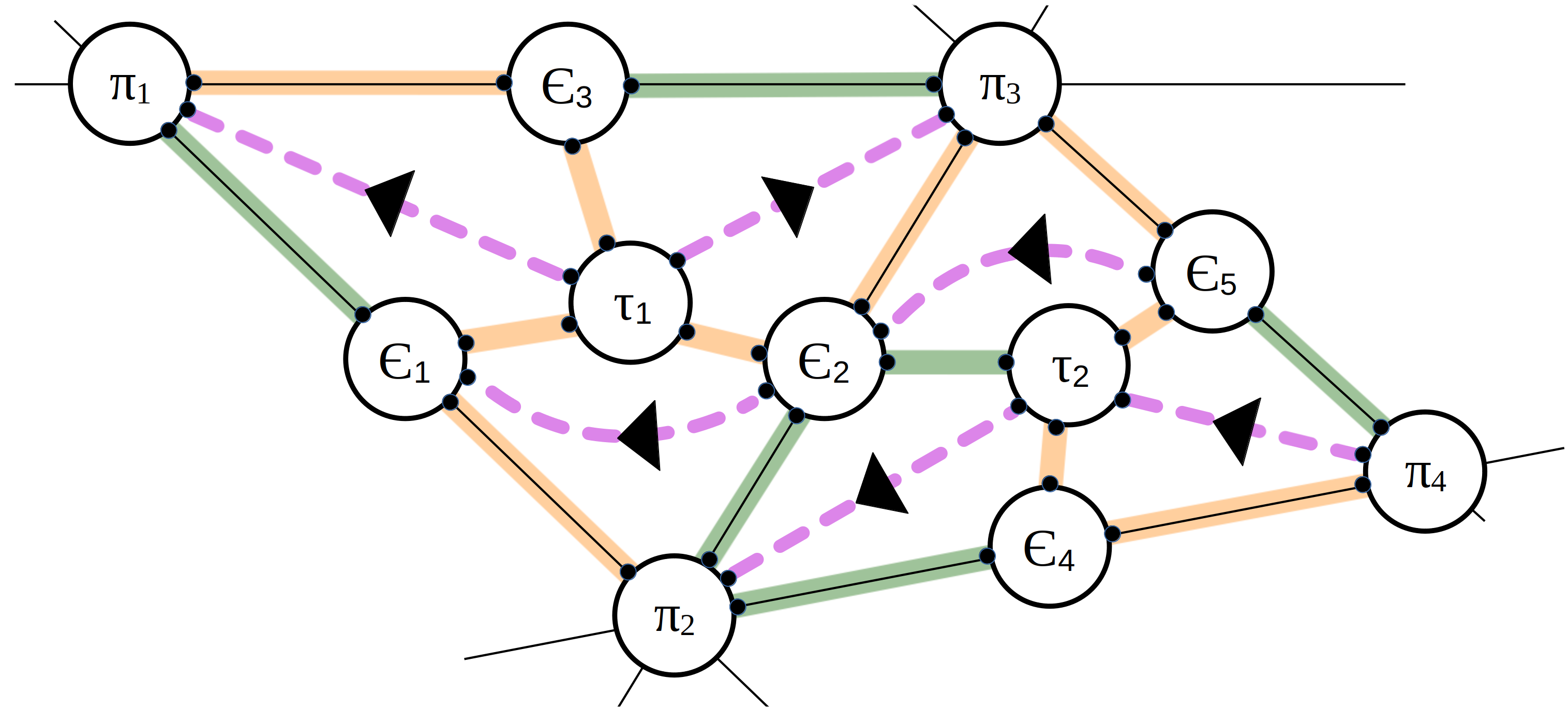}
    \caption{The physical couplings, as listed in Fig.~\ref{Fig:CR}, needed to implement the operator $B+B^\dagger$ (continuous thick lines), and operator $S$ (oriented dashed lines), for the data from Fig.~\ref{Fig:TM}.}
    \label{Fig:TCP}
\end{figure}

We now supply the values of the coupling coefficients $w_{\xi,\xi'}$. Boundary maps are standard operations in combinatorial geometry and are central to  simplicial homology \cite{Supplemental}. They lift to our Hilbert spaces as linear operators
\begin{equation}\label{Eq:B}
    B|[v_0 \ldots v_p]\rangle = \sum\nolimits_{i=0}^p (-1)^i |[v_0 \ldots \hat v_i \ldots v_p]\rangle,
\end{equation}
where hat indicates omission of that specific vertex. For faces, if the edges' orientations coincide with those induced by the face, as for $\tau_1$ in Fig.~\ref{Fig:TM}, then no negative signs occur in Eq.~\eqref{Eq:B}. However, this cannot be enforced throughout the triangulation and especially for neighboring facets. For example, negative signs do occur for $\tau_2$ in Fig.~\ref{Fig:TM}, for which $B|\tau_2\rangle = -|\epsilon_2\rangle +|\epsilon_4\rangle + |\epsilon_5\rangle$. For edges, the $B$-action takes $|[v_0,v_1]\rangle$ into $|[v_1]\rangle -|[v_0]\rangle$, and all vertices are sent to zero by $B$.

The matrix elements $\langle \xi|B|\xi'\rangle$ supply the strengths and signs of the couplings needed to implement the self-adjoint operator $B+B^\dagger$ with an actual metamaterial. They are exemplified in Fig.~\ref{Fig:TCP} for the geometry and resonators from Fig.~\ref{Fig:TM}. These couplings engage only nearest-neighboring resonators and all have strengths equal to one \cite{Footnote5}, though there are variations in the signs of the couplings (which pose no practical problem, see Fig.~\ref{Fig:CR}).

A $q$-cochain $\psi_q$ is just a map from $q$-simplices to $\CM$. An example relevant to us is $\delta_{\xi}(\xi')= \langle \xi' |\xi\rangle$, where $\xi$ and $\xi'$ are $q$-simplices. Up to a sign determined by the orientations of the simplices, its result is just $\delta_{\xi,\xi'}$. The cap product in combinatorial geometry is defined as
\begin{equation}
    \psi_q \smallfrown [v_0, \ldots, v_{q+p}] : = \psi_q([v_p,\ldots v_{q+p}])\, [v_0,\ldots, v_p].
\end{equation}
It was observed in \cite{HigsonKTII2004} that $\smallfrown$ and the fundamental cycle $|X\rangle=\sum_\tau |\tau\rangle$, engaging all (ordered) 2-cycles of a triangulation, can be combined into a linear operator on our Hilbert space $\Hh$, acting on the basis as
\begin{equation}
P|\xi\rangle : = \delta_{\xi} \smallfrown |X \rangle = \sum\nolimits_\tau \delta_{\xi} \smallfrown |\tau \rangle.
\end{equation}
This was used in \cite{HigsonKTII2004} to construct the linear operator 
\begin{equation}\label{Eq:T}
    T |\psi\rangle = \tfrac{1}{2}(P^\dagger + (-1)^{p} P)|\psi\rangle, \ |\psi\rangle \in \Hh_p, \ p=0,1,2,
\end{equation}
which implements the Poincar\'e duality isomorphism between the homologies of the dual Hilbert complexes 
\begin{equation}
    \Hh_0 \stackrel{\ B}{\leftarrow} \Hh_1 \stackrel{\ B}{\leftarrow} \Hh_2 \ \ {\rm and} \ \ \Hh_2 \stackrel{\ \ B^\dagger}{\leftarrow} \Hh_1 \stackrel{\ \ B^\dagger}{\leftarrow} \Hh_0.
\end{equation}
Note that $\Hh_{p=0,2}$ are not invariant under $T$; rather they are exchanged by $T$. Lastly, $T$ can be transformed into a self-adjoint operator $S=\imath^{p(p-1)+1}T$, which has the key property $BS + SB^\dagger=0$, assuring us that the Hamiltonians
\begin{equation}
H_\pm = B + B^\dagger \pm S
\end{equation}
have spectral gaps at zero \cite{HigsonKTI2004}. These are the two operators carrying Chern numbers $\pm 1$.

\begin{figure}
    \centering   \includegraphics[width=1.0\linewidth]{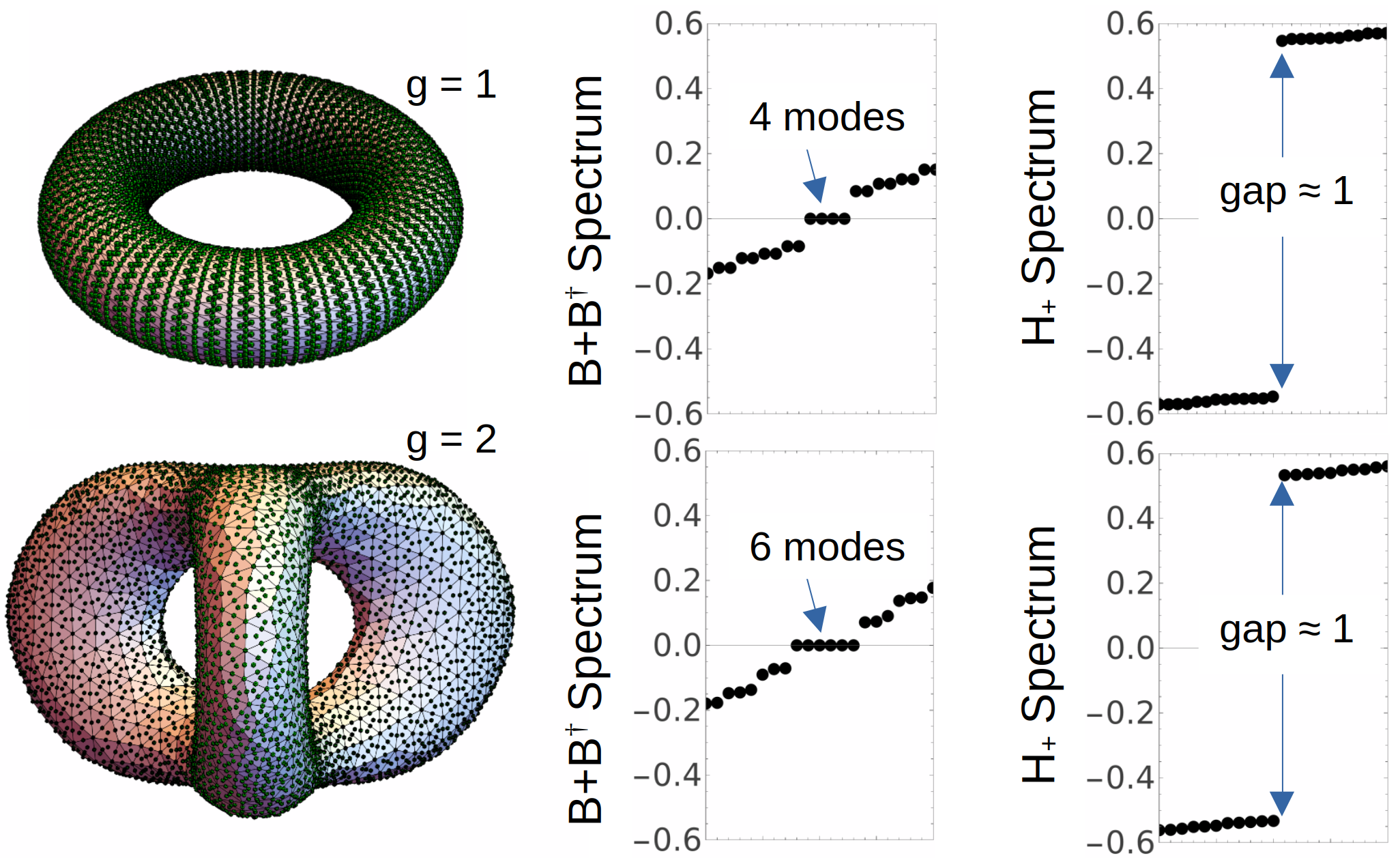}
    \caption{The bulk spectra of the indicated operators when deployed on surfaces of different genus $g$. The number of zero modes of $B+B^\dagger$ confirms the expected relation. The horizontal axes of the graphs represent the eigenvalue indices. The resonators are placed according to the seen triangulation.}
    \label{Fig:BS}
\end{figure}

The matrix elements $\langle \xi|S|\xi'\rangle$ are all pure imaginary and of magnitude $\tfrac{1}{2}$. They are exemplified in Fig.~\ref{Fig:TCP} for the geometry from Fig.~\ref{Fig:TM}. As advised in \cite{BarlasPRB2018}, we can promote $\sqrt{-1}$ to the matrix ${\scriptsize \begin{pmatrix} 0 & 1 \\ -1 & 0\end{pmatrix}}$ and $1$ to the $2 \times 2$ identity matrix, a move that removes the complex couplings at the expense of doubling the number of resonators (see Fig.~\ref{Fig:CR}(c)). The amplified Hilbert space $\CM ^2 \otimes \Hh$ comes equipped with a built-in symmetry, called $U$ in \cite{BarlasPRB2018}, and, if we decompose the Hilbert space according to this symmetry, then the amplification of $H_+$ is just ${\rm Diag}( H_+,H_-)$. Thus, as promised, we delivered an implementation of both topological models $H_\pm$ with one metamaterial using only nearest-neighbor positive and negative couplings of strengths $1$ or $\tfrac{1}{2}$ \cite{Footnote6}, in such a manner that their dynamics decouple and can be investigated independently (see \cite{WuSciBulletin2004} for further details).

{\it Detecting the topology of the surface.} Connected orientable surfaces are classified by their genus and, for a  genus $g$ surface $X$, the simplicial homology groups are $H_0(X)\simeq H_2(X)\simeq \ZM$ and $H_1(X)\simeq \ZM^{2g}$.  The zero modes of $B+ B^\dagger$ coincide with the generators of the homology groups. Thus, the number of its zero modes is $2 + 2g$ and, as exemplified in Fig.~\ref{Fig:BS}, the operator $B+B^\dagger$ can be used to identify the topology of a surface when one is handed a file with one of its triangulations. This principle can be very useful in situations where the surface is very complex or it is folded many times and a global visualization is impossible. While this is not new as a mathematical subject \cite{KaczynskiBook}, the interesting fact here is that we can actually ``hear" the topology of the surface if we coat it with a metamaterial based on the $B$-connections from Fig.~\ref{Fig:TCP}.

The zero modes of $B+B^\dagger$ are spectrally separated only for finite triangulations. Indeed, as the triangulations are sequentially refined, the discrete spectra seen in Fig.~\ref{Fig:BS} become gapless. However, when we add the operator $S$, a clean spectral gap of approximate size one opens in the middle of the spectra, and these gaps are stable against the refinements of the triangulations.

\begin{figure}
    \centering   \includegraphics[width=1.0\linewidth]{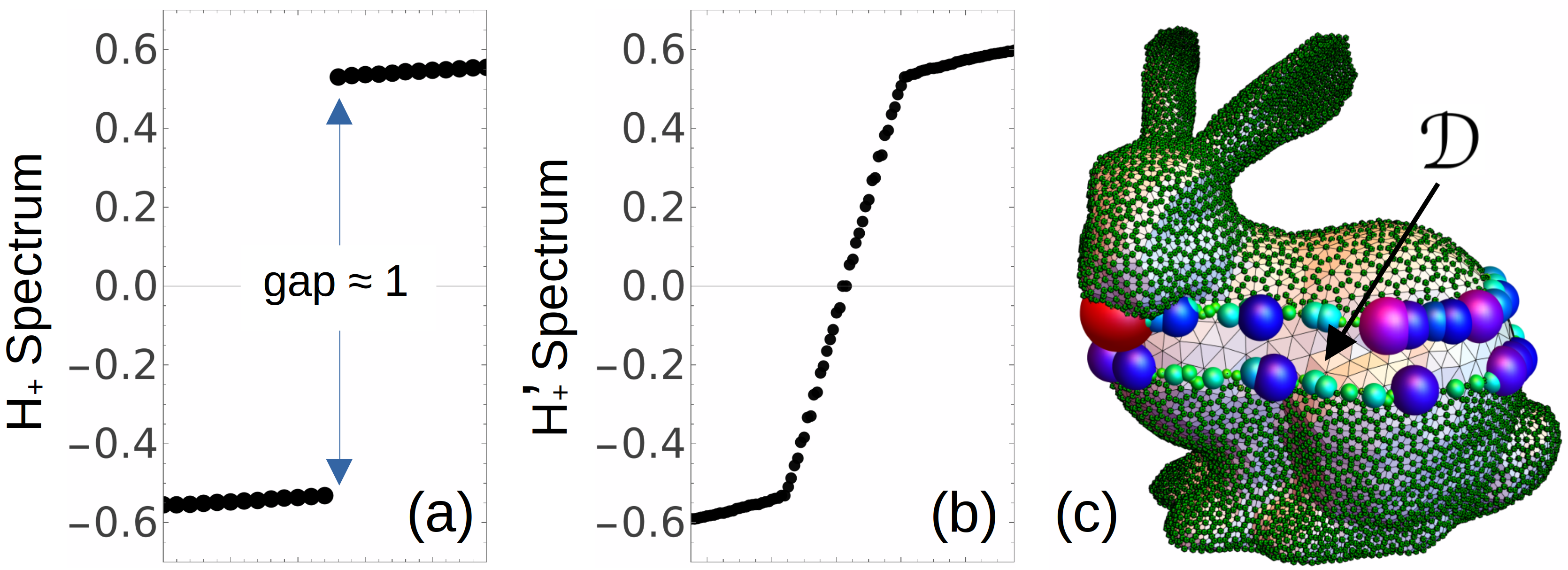}
    \caption{(a) Bulk spectrum when the model is deployed on a triangulation of the Stanford Bunny. (b) The spectrum of the model when the resonators inside the domain $\Dd$ are jammed. (c) The site amplitudes, encoded in the colors and radii of the balls, of a Gaussian packet $\sum_n e^{-\epsilon^{'2}_n/\sigma^2} \varphi'_n$, $\sigma = 0.6$, centered in the middle of the topological spectrum. Here, $\{\epsilon'_n, \varphi'_n\}$ is the eigensystem of $H'_+=H_+ + R_\Dd$.}
    \label{Fig:ES}
\end{figure}

{\it Topological edge modes.} We now implement $H_+$ on the triangulation of the real-object from Fig.~\ref{Fig:TE}, and confirm in Fig.~\ref{Fig:ES}(a) that the bulk spectrum is again gapped and free of defect-modes. The size of the gap is again approximately one. Next, we add the uniform onsite potential $R_\Dd=10^6 \sum_{\xi \in \Dd} |\xi\rangle \langle \xi|$ inside the domain $\Dd$ seen in Fig~\ref{Fig:ES}(c), which raises the resonant frequencies ({\it e.g.} by flooding the acoustic cavities with water) and effectively suppresses the dynamics at the operational frequencies \cite{Footnote7}. In Fig.~\ref{Fig:ES}(b), we confirm that this action fills the entire bulk spectral gap with spectrum, which is a hallmark of Chern models. Furthermore, in Fig~\ref{Fig:ES}(c), a sampling of that spectrum reveals that the associated modes live at the boundary $\partial \Dd$ of the region $\Dd$.

\begin{figure}
    \centering   \includegraphics[width=1.0\linewidth]{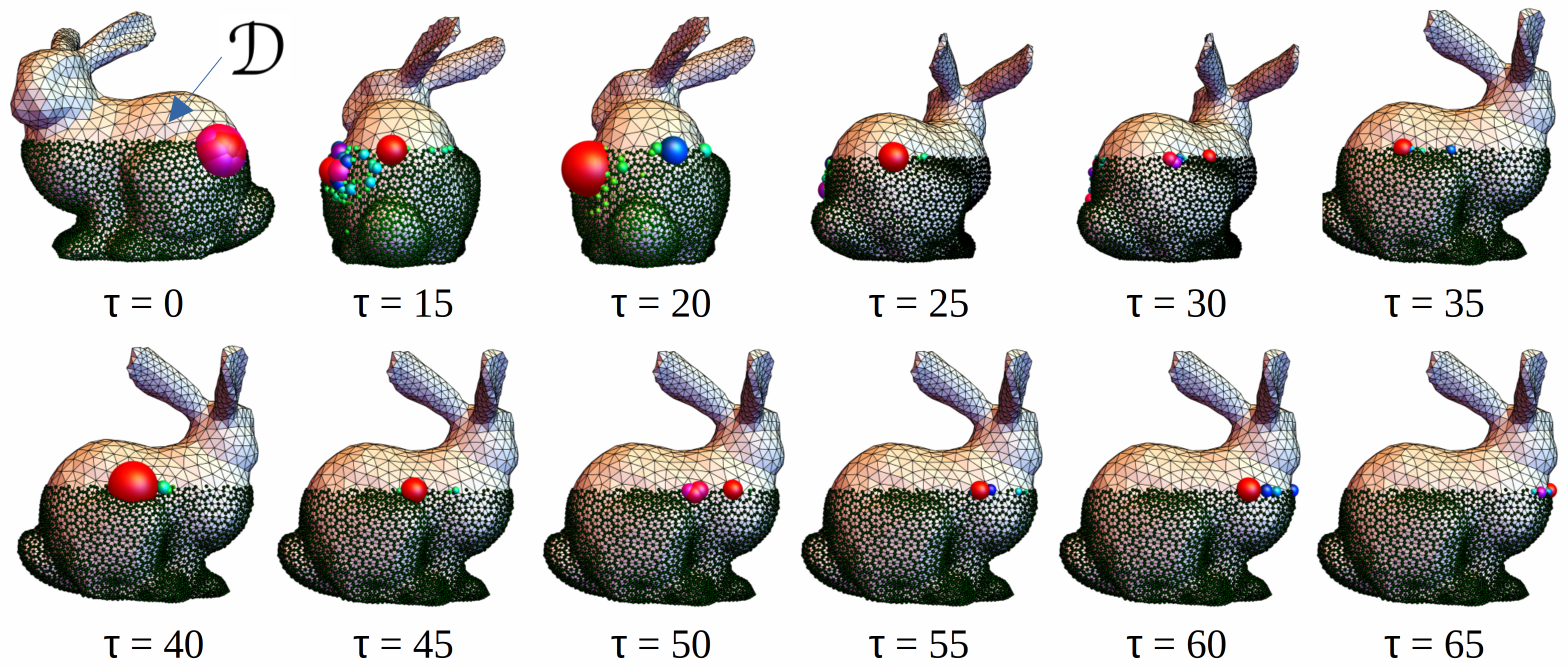}
    \caption{Time evolution of the initial configuration seen in the frame $t=0$, under the dynamics \eqref{Eq:TEvol}generated by $H'_+$.}
    \label{Fig:ED}
\end{figure}

In Fig~\ref{Fig:ED}, we chose $\Dd$ to be the whole upper half-side of the object, and we consider an initial state $\psi(0)$ localized at the boundary $\partial \Dd$ (see first frame in Fig.~\ref{Fig:ED}). Its time evolution under the dynamics generated by $H'_+=H_+ + R_\Dd$ is sampled in the figure, confirming that the excitation sets in motion one and only one wave-packet which travels only in one direction along $\partial \Dd$. If we replace $H_+$ by $H_-$, we have verified that there is again one and only one wave-packet which travels in the opposite direction. This confirms once again that the models $H_\pm$ carry Chern numbers $\pm 1$. Another feature revealed by the simulations is the non-dispersive nature of topological modes, much desired in practical applications. Indeed, as the wave-packet travels in Fig.~\ref{Fig:ED}, it stays concentrated around its center of mass, as opposed to slowly flattening and spreading along the boundary. The time-evolution was computed via exact diagonalization
\begin{equation}\label{Eq:TEvol}
    |\psi(\tau)\rangle = \sum\nolimits_n e^{\imath \epsilon'_n \tau} c_n |\varphi'_n\rangle, \quad H'_+|\varphi'_n\rangle=\epsilon'_n |\varphi'_n\rangle,
\end{equation}
and the $c_n$ coefficients were extracted from the expansion $|\psi(0)\rangle = \sum_n c_n |\varphi'_n\rangle$ of the initial state \cite{Footnote8}.

In conclusion, we devised an algorithm which takes the triangulation of a closed orientable surface at the input and coats the surface with a metamaterial whose dynamics displays a clean topological spectral gap carrying a non-trivial Chern number. The algorithm uses the matrix elements of the boundary and Poincar\'e duality operators, which are available in universal forms for any Hilbert complex derived from triangulations of closed oriented surfaces. We confirmed that, after removing or jamming the resonators in a region of the surface, one-way propagation channels develop along the boundary of the region. Our findings represent an important step towards deploying topological metamaterials on the surfaces of real-world objects, hence bringing the topological metamaterials closer to real-world applications.

\begin{acknowledgments}
{\it Acknowledgments.} This work was supported by the U.S. National Science Foundation through grants CMMI-2131760 and DMS-1952669, and from the U.S. Army Research Office through contract W911NF-23-1-0127.
\end{acknowledgments}


\begin{thebibliography}{1}
\bibitem{Footnote1}  A refinement will  be followed by a scaling of the surface such that the scale of the triangles remain unchanged.
\bibitem{Footnote2} A metamaterial is a collection of coupled classical resonators or artificial atoms, each carrying a finite number of modes/orbitals. As opposed to a natural material, in a metamaterial one has full control over the internal structure of the resonators/artificial atoms and over the strengths/phases of the couplings.
\bibitem{Footnote3} By definition, when a one-way topological wave-channel is excited by a point source, material deformations propagate in only one direction along the channel, and this propagation cannot be stopped by material imperfections.
\bibitem{Supplemental4} The self-oscillating air pressure is $p(x,t)=p(x)e^{\imath \omega_0 t}$, with amplitude $p(x)$ constant along the cross section and, along the acoustic cavity, very similar to the amplitude of the first resonant mode of a free-ends rope under tension.
\bibitem{Footnote5} Note that the energy scale is set by the resonant frequency $\omega_0$ of the uncoupled resonators.
\bibitem{Footnote6} If desired, the coupling strengths can be forced to be one, by replacing $H_+$ with $B+B^\dagger + 2 S$.
\bibitem{Footnote7} In practical applications, the wave-channels need to be reconfigurable and, as such, removal of the resonators is not an option. Suppressing dynamics as explained can be done fast and reversibly, hence it is the preferred option.
\bibitem{Footnote8} For the classical metamaterial, the eigenmodes evolve in time as $\varphi'_n(t)= e^{\imath \omega'_n t}\varphi'_n$ rather than $\varphi'_n(t)= e^{\imath \epsilon'_n t}\varphi'_n$, but the expansion $\omega'_n=\omega_0\sqrt{1+\epsilon'_n/\omega^2_0} \approx \omega_0+\tfrac{1}{2}\epsilon'_n/\omega_0 $, $\epsilon'_n=\omega_n^{'2}-\omega_0^2$, justifies \eqref{Eq:TEvol} if $\tau$ there stands for $t/2\omega_0$. The factor $e^{\imath \omega_0 t}$ does not affect the plotted amplitudes. This is standard in the coupled-mode theory, which is valid in the regime $\epsilon/\omega_0^2\ll 1$.
\end{thebibliography}
\end{document}